\documentclass[useAMS,onecolumn,usenatbib]{mn2e}

\usepackage{graphicx}

\newcommand{\prt}{\partial}

\title[Multi-transonic pseudo-Schwarzschild accretion]
{Critical properties and stability of stationary solutions  
in multi-transonic pseudo-Schwarzschild accretion }
\author[Chaudhury et al.]
{Soumini Chaudhury,$^{1,2}$\thanks{soumini.chaudhury@saha.ac.in}
Arnab K. Ray$^{3,4}$\thanks{akr@iucaa.ernet.in}
and Tapas K. Das$^{3,5}$\thanks{tapas@mri.ernet.in}\\
$^{1}$Department of Physics, Jadavpur University, 
Jadavpur, Kolkata 700032, India\\
$^{2}$Saha Institute of Nuclear Physics, Sector 1, Block AF,
Bidhannagar, Kolkata 700064, India\\
$^{3}$Harish--Chandra Research Institute, Chhatnag Road, Jhunsi,
Allahabad 211019, India\\
$^{4}$Inter--University Centre for Astronomy and Astrophysics, Post
Bag 4, Ganeshkhind, Pune University Campus, Pune 411007, India\\
$^{5}$Theoretical Institute for Advanced Research in Astrophysics,
101, Section 2, Kuang Fu Road, Hsinchu, Taiwan} 

\begin{document}



\maketitle

\label{firstpage}

\begin{abstract}
For inviscid, rotational accretion flows, both isothermal and
polytropic, a simple dynamical systems analysis of the critical
points has given a very accurate mathematical scheme to understand
the nature of these points, for {\em any} pseudo-potential by 
which the flow may be driven on to a Schwarzschild black hole. 
This allows for a complete classification of the critical 
points for a wide range of flow parameters, and shows that the 
only possible critical points for this kind of flow are saddle
points and centre-type points. A restrictive upper bound on the 
angular momentum of critical solutions has been established. 
A time-dependent perturbative study reveals that 
the form of the perturbation equation, for both isothermal and 
polytropic flows, is invariant under the choice of any particular 
pseudo-potential. Under generically true outer boundary conditions, 
the inviscid flow has been shown to be stable under 
an adiabatic and radially propagating perturbation. 
The perturbation equation has also served the dual purpose
of enabling an understanding of the acoustic geometry for inviscid
and rotational flows.
\end{abstract}


\section{Introduction}
\label{sec1}

Multi-transonicity in axisymmetric accretion becomes possible for a
certain range of values of the specific angular momentum of the flow,
when the number of critical points in the accreting system exceeds
one, unlike in the case of spherically symmetric accretion. 
A substantial body of work over the past many years, has argued 
well the case for multi-transonicity in axisymmetric
accretion~\citep{az81,fuk87,skc90,ky94,yk95,par96,la97,lyyy97,pa97,
das02,bdw04,das04,abd06,dbd06}. 
Transonicity would imply that the bulk velocity of the 
flow would be matched by the speed of acoustic propagation in the 
accreting fluid. In this situation a subsonic-to-supersonic 
transition or vice versa can take place in the flow either 
continuously or discontinuously. In the former case the flow is 
smooth and regular through a critical point for transonicity 
(more specially, this can be a sonic point, at which the bulk flow
exactly matches the speed of sound), while in the latter case, 
there will arise a shock~\citep{c89}. 
The possibility of both kinds of transition 
happening in an accreting system is very much real, and much 
effort has been made so far in studying these phenomena. 
For accretion on to a black hole especially, the argument that 
the inner boundary condition at the event horizon will lead to
the exhibition of transonic properties in the flow, has been 
established well~\citep{skc90}. However, for all the importance 
of transonic flows,
there exists as yet no general mathematical prescription that 
allows for a direct understanding of the nature of the critical
points of the flow (and the physical solutions which pass
through them) without having to take recourse to the conventional
approach of numerically integrating the governing non-linear flow 
equations, and then looking for any physically meaningful transonic 
behaviour under a given set of flow parameters. 

This paper purports to address that particular issue. To do so
it has been necessary to adopt mathematical methods from the 
study of dynamical systems~\citep{js99}. In accretion literature one 
would come across some relatively recent works which had indeed 
made use of the techniques of dynamical systems~\citep{rb02,ap03}, 
but the present treatment, it is being 
claimed, is much broader and more comprehensive in its scope and
objectives, than those reported previously. Through this approach 
a complete and mathematically rigorous prescription has been made 
for the exclusive nature of the critical points in axisymmetric 
pseudo-Schwarzschild accretion, and, in consequence, the possible 
behaviour of the flow solutions passing through those points 
and their immediate neighbourhood. In conjunction with a knowledge 
of the boundary conditions for physically feasible inflow solutions, 
this makes it possible to form an immediate qualitative notion 
of the solution topologies. What is more, this study has been 
shown to set a more restrictive condition on sub-Keplerian flows 
passing through the critical points. 

Beyond this stage, time-dependence has been brought into the 
flow equations. This has been easy because, although this study
is devoted to black hole accretion, the Newtonian construct of space 
and time has been preserved, with the general relativistic
effects of strong gravity in the vicinity of a Schwarzschild 
black hole having been represented through a modified 
pseudo-Newtonian potential, for which, various prescriptions 
exist in accretion literature~\citep{pw80,nw91,abn96}. 
Thus it has been possible in this 
Newtonian framework, to carry out a time-dependent stability 
analysis of the physically well-behaved 
and relevant stationary solutions in a 
straightforward manner. The stability of these solutions
has been shown to be dependent on the asymptotic conditions on
large length scales of the accretion disc, and since these 
conditions are the same for the choice of any
pseudo-Newtonian potential, it has been argued that the flow
solutions will remain stable under linearised perturbations in
all cases. Another interesting aspect of this perturbative study 
in real time has been that the equation of motion for the 
propagation of the perturbation yields a metric that is entirely
identical to the metric of an analog acoustic black hole obtained
through very different means~\citep{vis98}. 

One important feature of the mathematical treatment in this paper
is worth stressing. The entire study --- on the nature of the critical
points as well as on the time-dependent stability of the flow --- 
has been carried out in terms of a general potential only. Therefore,
the mathematical results will hold equally well under the choice of 
{\em any} particular potential that will drive the flow --- both 
isothermal as well as a general polytropic --- on to the black hole. 

Finally, the range of applicability of the mathematical methods of 
dynamical systems goes much beyond a gravitational system defined 
purely within the Newtonian framework. In a rigorously general 
relativistic stationary flow, described variously by either the 
Schwarzschild metric or the Kerr metric, it is eminently possible 
to carry out a mathematical analysis in the same principle in which 
it has been presented in this work. These general relativistic cases 
will be reported separately, while the present work may be considered 
to be the first in a series in which 
the usefulness of a dynamical systems approach in astrophysical
problems has been considered at length. 

\section{The Equations of the flow and its fixed points}
\label{sec2}

It is a standard practice to consider a thin, rotating, axisymmetric,
inviscid steady flow, with the condition of hydrostatic equilibrium
imposed along the transverse direction~\citep{mkfo84,fkr02}. The two 
equations which determine the drift in the radial direction are 
Euler's equation,
\begin{equation}
\label{euler}
v \frac{\mathrm{d}v}{\mathrm{d}r} 
+ \frac{1}{\rho}\frac{\mathrm{d}P}{\mathrm{d}r} 
+ \phi^{\prime}(r) - \frac{\lambda^2}{r^3} = 0 
\end{equation}
and the equation of continuity,
\begin{equation}
\label{con}
\frac{\mathrm{d}}{\mathrm{d}r}\left(\rho vrH \right) = 0 
\end{equation}
in which, $\phi(r)$ is the generalised pseudo-Newtonian potential
driving the flow (with the prime denoting a spatial derivative), 
$\lambda$ is the conserved angular momentum of the flow, $P$ is 
the pressure of the flowing gas and $H \equiv H(r)$ is the local 
thickness of the disc~\citep{fkr02}, respectively. 

The pressure, $P$, is prescribed by an equation of state for the 
flow~\citep{sc39}. 
As a general polytropic it is given as $P=K \rho^{\gamma}$,
while for an isothermal flow the pressure is given by
$P= \rho {\kappa}T/\mu m_{\mathrm{H}}$, in all of which, 
$K$ is a measure of the entropy in the flow, $\gamma$ is the 
polytropic exponent, $\kappa$ is Boltzmann's constant, $T$ is the 
constant temperature, $m_{\mathrm{H}}$ is the mass of a hydrogen
atom and $\mu$ is the reduced mass, respectively. The 
function $H$ in Eq.(\ref{con}) will be determined according to the 
way $P$ has been prescribed~\citep{fkr02}, while transonicity in the 
flow will be measured by scaling the bulk velocity of the flow 
with the help of the local speed of sound, given as 
$c_{\mathrm{s}} = (\partial P/\partial \rho)^{1/2}$. In what follows, 
the flow properties will be taken up separately for the two cases, 
i.e. polytropic and isothermal. 

\subsection{Polytropic flows}

With the polytropic relation specified for $P$, it is a 
straightforward exercise to set down in terms of the speed of 
sound, $c_{\mathrm{s}}$, a first integral of Eq.(\ref{euler}) as, 
\begin{equation}
\label{eupol1st}
\frac{v^2}{2} + n c_{\mathrm{s}}^2 + \phi (r) 
+ \frac{\lambda^2}{2 r^2} = \mathcal{E} 
\end{equation}
in which $n=(\gamma -1)^{-1}$ and the integration constant 
$\mathcal{E}$ is the Bernoulli constant. Before moving on to 
find the first integral of Eq.(\ref{con}) it should be important
to derive the functional form of $H$. Assumption of hydrostatic
equilibrium in the vertical direction deliver this form to be 
\begin{equation}
\label{aitchpol}
H = c_{\mathrm{s}} \left(\frac{r}{\gamma \phi^{\prime}}\right)^{1/2}
\end{equation}
with the help of which, the first integral of Eq.(\ref{con}) could
be recast as 
\begin{equation}
\label{conpol1st}
c_{\mathrm{s}}^{2(2n +1)} \frac{v^2 r^3}{\phi^{\prime}} 
= \frac{\gamma}{4 \pi^2} \dot{\mathcal{M}}^2 
\end{equation}
where $\dot{\mathcal{M}} = (\gamma K)^n \dot{m}$~\citep{skc90} 
with $\dot{m}$, an integration constant itself, being physically the 
matter flow rate. 

To obtain the critical points of the flow, it should be necessary
first to differentiate both Eqs.(\ref{eupol1st}) and (\ref{conpol1st}),
and then, on combining the two resultant expressions, to arrive at 
\begin{equation}
\label{dvdrpol}
\left(v^2 - \beta^2 c_{\mathrm{s}}^2 \right)
\frac{\mathrm{d}}{\mathrm{d}r}(v^2) = \frac{2 v^2}{r}
\left[ \frac{\lambda^2}{r^2} - r \phi^{\prime} 
+ \frac{1}{2}\beta^2 c_{\mathrm{s}}^2 
\left(3 - r \frac{\phi^{\prime \prime}}{\phi^{\prime}} \right) \right ]
\end{equation}
with $\beta^2 = 2(\gamma +1)^{-1}$. The critical points of the flow
will be given by the condition that the entire right hand side of 
Eq.(\ref{dvdrpol}) will vanish along with the coefficient of 
${\mathrm{d}}(v^2)/{\mathrm{d}r}$. Explicitly written down,
following some rearrangement of terms, this will give the two
critical point conditions as, 
\begin{equation}
\label{critconpol}
v_{\mathrm{c}}^2 = \beta^2 c_{\mathrm{sc}}^2
= 2\left[r_{\mathrm{c}} \phi^{\prime}(r_{\mathrm{c}}) 
- \frac{\lambda^2}{r_{\mathrm{c}}^2} \right] \left[3 - r_{\mathrm{c}}
\frac{\phi^{\prime \prime}(r_{\mathrm{c}})}{\phi^{\prime}(r_{\mathrm{c}})} 
\right]^{-1} 
\end{equation}
with the subscript ${\mathrm{c}}$ labelling critical point values. 

To fix the critical point coordinates, $v_{\mathrm{c}}$ and $r_{\mathrm{c}}$,
in terms of the system constants, one would
have to make use of the conditions given by Eqs.(\ref{critconpol}) along
with Eq.(\ref{eupol1st}), to obtain 
\begin{equation}
\label{efixcrit}
\frac{2 \gamma}{\gamma -1} 
\left[r_{\mathrm{c}} \phi^{\prime}(r_{\mathrm{c}})
- \frac{\lambda^2}{r_{\mathrm{c}}^2} \right] \left[3 - r_{\mathrm{c}}
\frac{\phi^{\prime \prime}(r_{\mathrm{c}})}{\phi^{\prime}(r_{\mathrm{c}})}
\right]^{-1} + \phi (r_{\mathrm{c}}) 
+ \frac{\lambda^2}{2 r_{\mathrm{c}}^2} = \mathcal{E} 
\end{equation}
from which it is easy to see that solutions of $r_{\mathrm{c}}$ may
be obtained in terms of $\lambda$ and $\mathcal{E}$ only, i.e. 
$r_{\mathrm{c}}=f_1(\lambda, \mathcal{E})$. 
Alternatively, $r_{\mathrm{c}}$
could be fixed in terms of $\lambda$ and $\dot{\mathcal{M}}$. By making
use of the critical point conditions in Eq.(\ref{conpol1st}) one could 
write 
\begin{equation}
\label{dotmfix}
\frac{4 \pi^2 \beta^2 r_{\mathrm{c}}^3}
{\gamma\phi^{\prime}(r_{\mathrm{c}})}
\left( \frac{2}{\beta^2}
\left[r_{\mathrm{c}} \phi^{\prime}(r_{\mathrm{c}})
- \frac{\lambda^2}{r_{\mathrm{c}}^2} \right] \left[3 - r_{\mathrm{c}}
\frac{\phi^{\prime \prime}(r_{\mathrm{c}})}{\phi^{\prime}(r_{\mathrm{c}})}
\right]^{-1}\right)^{2(n +1)} = {\dot{\mathcal{M}}}^2  
\end{equation}
with the obvious implication being that the dependence of $r_{\mathrm{c}}$
will be given as $r_{\mathrm{c}}= f_2(\lambda, \dot{\mathcal{M}})$. 
Comparing these two alternative means of fixing $r_{\mathrm{c}}$, the 
next logical step would be to say that for the fixed points, and for 
the solutions passing through them, it should suffice to specify either 
$\mathcal{E}$ or $\dot{\mathcal{M}}$~\citep{skc90}. 

\subsection{Isothermal flows}

For isothermal flows, one has to go back to Eq.(\ref{euler}) and use
the linear dependence between $P$ and $\rho$ as the appropriate equation
of state. On doing so, the first integral of Eq.(\ref{euler}) is given as
\begin{equation}
\label{euiso1st}
\frac{v^2}{2} + c_{\mathrm s}^2 \ln \rho + \phi (r)
+ \frac{\lambda^2}{2 r^2} = \mathcal{C}
\end{equation}
with $\mathcal{C}$ being a constant of integration. For flow solutions
which specifically decay out to zero at very large distances, the constant 
$\mathcal{C}$ can be determined in terms of the ``ambient conditions" as
$\mathcal{C} = c_{\mathrm s}^2 \ln \rho_\infty$. The thickness
of the disc can be shown to bear the dependence
\begin{equation}
\label{aitchiso}  
H = c_{\mathrm{s}} \left(\frac{r}{\phi^{\prime}}\right)^{1/2}
\end{equation}
with the help of which, the first integral of the continuity equation
is given as, 
\begin{equation}
\label{coniso1st}
\frac{\rho^2 v^2 r^3}{\phi^{\prime}}
= \frac{\dot{m}^2}{4 \pi^2 c_{\mathrm{s}}^2} 
\end{equation}
As has been done for polytropic flows, 
both Eqs.(\ref{euiso1st}) and (\ref{coniso1st}) are 
to be differentiated and the results combined to give,
\begin{equation}
\label{dvdriso}
\left(v^2 -  c_{\mathrm{s}}^2 \right)
\frac{\mathrm{d}}{\mathrm{d}r}(v^2) = \frac{2 v^2}{r}
\left[ \frac{\lambda^2}{r^2} - r \phi^{\prime} 
+ \frac{1}{2} c_{\mathrm{s}}^2
\left(3 - r \frac{\phi^{\prime \prime}}{\phi^{\prime}} \right) \right ]
\end{equation}
from which the critical point conditions are easily identified as,
\begin{equation}
\label{critconiso}
v_{\mathrm{c}}^2 = c_{\mathrm{s}}^2
= 2\left[r_{\mathrm{c}} \phi^{\prime}(r_{\mathrm{c}})
- \frac{\lambda^2}{r_{\mathrm{c}}^2} \right] \left[3 - r_{\mathrm{c}}
\frac{\phi^{\prime \prime}(r_{\mathrm{c}})}{\phi^{\prime}(r_{\mathrm{c}})}
\right]^{-1}
\end{equation}
In this isothermal system, the speed of sound, $c_{\mathrm{s}}$, is 
globally a constant, and so having arrived at the critical point
conditions, it should be easy to see that $r_{\mathrm{c}}$ and 
$v_{\mathrm{c}}$ have already been fixed in terms of a global 
constant of the system. The speed of sound can further be written
in terms of the temperature of the system as 
$c_{\mathrm{s}} = \Theta T^{1/2}$, where 
$\Theta = (\kappa/\mu m_{\mathrm H})^{1/2}$, and, therefore, 
it should be entirely possible to give a functional dependence for
$r_{\mathrm{c}}$, as $r_{\mathrm{c}} = f_3(\lambda, T)$. 

\section{Nature of the fixed points : A dynamical systems study}
\label{sec3}

The equations governing the flow in an accreting system are in 
general first-order non-linear differential equations. There is
no standard prescription for a rigorous mathematical analysis of
these equations. Therefore, for any understanding of the behaviour
of the flow solutions, a numerical integration is in most cases 
the only recourse. On the other hand, an alternative approach 
could be made to this question, if the governing equations are
set up to form a standard first-order dynamical system~\citep{js99}. 
This is a very usual practice in general fluid dynamical 
studies~\citep{bdp93}, 
and short of carrying out any numerical integration, this approach
allows for gaining physical insight into the behaviour of the 
flows to a surprising extent. As a first step towards this end,
for the stationary polytropic flow, as given by Eq.(\ref{dvdrpol}),
it should be necessary to parametrise this equation and set up 
a coupled autonomous first-order dynamical system as~\citep{js99} 
\begin{eqnarray}
\label{dynsys}
\frac{\mathrm{d}}{\mathrm{d}\tau}(v^2)&=& 2v^2 \left[  
\frac{\lambda^2}{r^2} - r \phi^{\prime} + \frac{1}{2}
\beta^2 c_{\mathrm{s}}^2 \left( 3 - r \frac{\phi^{\prime \prime}}
{\phi^{\prime}} \right) \right] \nonumber \\
\frac{\mathrm{d}r}{\mathrm{d} \tau}&=& r \left(v^2 - 
\beta^2 c_{\mathrm{s}}^2 \right)
\end{eqnarray}
in which $\tau$ is an arbitrary mathematical parameter. With respect
to accretion studies in particular, this kind of parametrisation
has been reported before~\citep{rb02,ap03}, 
but here the possibility of a more thorough 
mathematical description of the nature of the critical points is 
being explored.  

The critical points have themselves been fixed in terms of the flow
constants. About these fixed point values, upon using a perturbation 
prescription of the kind 
$v^2 = v_{\mathrm{c}}^2 + \delta v^2$, $c_{\mathrm{s}}^2 = 
c_{\mathrm{sc}}^2 + \delta c_{\mathrm{s}}^2$ and 
$r = r_{\mathrm{c}} + \delta r$, one could derive a set of two 
autonomous first-order linear differential equations in the 
$\delta r$ --- $\delta v^2$ plane, with $\delta c_{\mathrm{s}}^2$ itself
having to be first expressed in terms of $\delta r$ and $\delta v^2$,
with the help of Eq.(\ref{conpol1st}) --- the continuity equation 
--- as 
\begin{equation}
\label{varsound}
\frac{\delta c_{\mathrm{s}}^2}{c_{\mathrm{sc}}^2} = - \frac{\gamma -1}
{\gamma + 1} \left( \frac{\delta v^2}{v_{\mathrm{c}}^2} 
+ \left[3 - r_{\mathrm{c}}\frac{\phi^{\prime \prime}(r_{\mathrm{c}})}
{\phi^{\prime}(r_{\mathrm{c}})} \right]
\frac{\delta r}{r_{\mathrm{c}}}
\right)
\end{equation}
The resulting coupled set of linear equations in $\delta r$ and
$\delta v^2$ will be given as 
\begin{eqnarray}
\label{lindynsys}
\frac{1}{{2v_{\mathrm{c}}^2}}\frac{\mathrm{d}}{\mathrm{d}\tau}
(\delta v^2) &=&  \frac{\mathcal A}{2}\left(\frac{\gamma -1}{\gamma + 1} 
\right) \delta v^2 - \left [\frac{2 \lambda^2}{r_{\mathrm{c}}^3} + 
\phi^{\prime}(r_{\mathrm{c}}) + r_{\mathrm{c}}\phi^{\prime \prime}
(r_{\mathrm{c}}) + \frac{\beta^2}{2} 
\frac{\phi^{\prime \prime}(r_{\mathrm{c}})}{\phi^{\prime}(r_{\mathrm{c}})}
{c_{\mathrm{sc}}^2} \mathcal{B} + \frac{\beta^2}{2} 
\left(\frac{\gamma -1}{\gamma + 1} \right)
\frac{{c_{\mathrm{sc}}^2}}{r_{\mathrm{c}}}
{\mathcal A}^2 \right] \delta r \nonumber \\
\frac{1}{r_{\mathrm{c}}}\frac{\mathrm{d}}{\mathrm{d}\tau}(\delta r)
&=& \frac{2\gamma}{\gamma + 1} \delta v^2 - \mathcal{A} \left(
\frac{\gamma -1}{\gamma + 1} \right)
\frac{{v_{\mathrm{c}}^2}}{r_{\mathrm{c}}} \delta r
\end{eqnarray}
in which 
\begin{displaymath}
\label{coeffs}
\mathcal{A} = r_{\mathrm{c}}\frac{\phi^{\prime \prime}(r_{\mathrm{c}})}
{\phi^{\prime}(r_{\mathrm{c}})} - 3 \,,  \qquad
\mathcal{B} = 1 + r_{\mathrm{c}}
\frac{\phi^{\prime \prime \prime}(r_{\mathrm{c}})}
{\phi^{\prime \prime}(r_{\mathrm{c}})}
- r_{\mathrm{c}}\frac{\phi^{\prime \prime}(r_{\mathrm{c}})}
{\phi^{\prime}(r_{\mathrm{c}})}
\end{displaymath}
Trying solutions of the kind $\delta v^2 \sim \exp(\Omega \tau)$
and $\delta r \sim \exp(\Omega \tau)$ in Eqs.(\ref{lindynsys}), will 
deliver the eigenvalues $\Omega$ --- growth rates of $\delta v^2$ and 
$\delta r$ --- as 
\begin{equation}
\label{eigen}
\Omega^2 = \frac{4 r_{\mathrm{c}}  
\phi^{\prime}(r_{\mathrm{c}})c_{\mathrm{sc}}^2}{(\gamma + 1)^2} 
\left( \left[ \left(\gamma - 1 \right)
{\mathcal A} - 2 \gamma \left(4 + {\mathcal A} \right) + 2 \gamma 
{\mathcal{B}}\left(1 + \frac{3}{\mathcal A}\right) \right] 
- \frac{\lambda^2}{\lambda_{\mathrm K}^2(r_{\mathrm{c}})} 
\left[4 \gamma + \left(\gamma - 1 \right){\mathcal A} + 2 \gamma 
{\mathcal{B}}\left(1 + \frac{3}{\mathcal A}\right) \right]
\right)
\end{equation}
where $\lambda_{\mathrm K}^2(r) = r^3 \phi^{\prime}(r)$. 

For isothermal flows, starting from Eq.(\ref{dvdriso}), a similar 
expression for the related eigenvalues may likewise be derived. The
algebra in this case is much simpler and it is an easy  
exercise to assure oneself that for isothermal flows one simply needs
to set $\gamma = 1$ in Eq.(\ref{eigen}), to arrive at a corresponding 
relation for $\Omega^2$. However, it should be incorrect to assume 
that in this kind of study, one could always treat isothermal flows 
simply as a special physical case of general polytopic flows. 
For polytropic flows the position of the fixed points, under a given
form of $\phi(r)$, will be determined either by Eq.(\ref{efixcrit}) 
or by Eq.(\ref{dotmfix}). For isothermal flows the fixed points are 
simply to be determined from the critical point conditions, 
as Eqs.(\ref{critconiso}) give them, 
since $c_{\mathrm{s}}$ is globally a constant in this case. 

Once the position of a critical point, $r_{\mathrm{c}}$, has been 
ascertained, it is then a straightforward task to find the nature 
of that critical point by using $r_{\mathrm{c}}$ in Eq.(\ref{eigen}). 
Since it has been discussed in Section~\ref{sec2} that $r_{\mathrm{c}}$
is a function of $\lambda$ and $T$ for isothermal flows, and a function
of $\lambda$ and $\mathcal E$ (or $\dot{\mathcal M}$) for polytropic
flows, it effectively implies that $\Omega^2$ can, in principle, be 
rendered as a function of the flow parameters for either kind of flow. 
A generic conclusion that can be drawn about the critical points from 
the form of $\Omega^2$ in Eq.(\ref{eigen}), is that {\em for a conserved 
pseudo-Schwarzschild axisymmetric flow driven by any potential, the only 
admissible critical points will be saddle points and centre-type points}. 
For a saddle point, $\Omega^2 > 0$, while for a centre-type point, 
$\Omega^2 < 0$. Once the behaviour of all the physically relevant 
critical points has been understood in this way, a complete qualitative
picture of the flow solutions passing through these points (if they 
are saddle points), or in the neighbourhood of these points (if they
are centre-type points), can be constructed, along with an impression
of the direction that these solutions can have in the phase portrait
of the flow~\citep{js99}. 

A further interesting point that can be appreciated from the derived 
form of $\Omega^2$, is related to the admissible range of values for 
a sub-Keplerian flow passing through a saddle point. It is self-evident
that for this kind of flow, 
$(\lambda/\lambda_{\mathrm K})^2 <1$~\citep{az81}. 
However, a look at Eq.(\ref{eigen}) will reveal that a more restrictive
upper bound on $\lambda/\lambda_{\mathrm K}$ can be imposed under the
requirement that $\Omega^2 > 0$ for a saddle point, and this restriction
will naturally be applicable to solutions which pass through such a 
point. This is entirely 
a physical conclusion, and yet its establishing has been achieved 
through a mathematical parametrisation of a dynamical system. 

\section{Numerical results} 
\label{sec4}

It has been argued with the help of Eq.(\ref{eigen}), that the spatial 
coordinate of the critical points, $r_{\mathrm{c}}$ (which in its turn 
depends on the flow parameters), will determine the eigenvalues and the 
corresponding nature of a given critical point. This will also
necessitate an explicitly given functional form of $\phi (r)$. 
A pseudo-Schwarzschild flow is driven by a pseudo-Newtonian potential,
which describes general relativistic effects which are most
important for accretion disc structures in the vicinity of a
Schwarzschild black hole. Introduction of such potentials will 
allow for the investigation of complicated physical processes 
taking place in disc accretion in a semi-Newtonian framework, by 
avoiding the difficulties of a purely general relativistic treatment. 
Through this approach most of the features of space-time around a 
compact object are retained and some crucial properties of analogous
relativistic solutions of disc structures can be reproduced with a
high degree of accuracy. 

In the present treatment, four such pseudo-Newtonian potentials have 
been chosen to make a graphical representation of both the qualitative
and the quantitative features of the critical points with respect to 
a given set of flow parameters ($\mathcal E$, $\gamma$ and $\lambda$ 
for polytropic flows, and $T$ and $\lambda$ for isothermal flows). 
In general a potential has been represented by $\phi_i (r)$, with
$\{i=1,2,3,4\}$. Written explicitly, each of these potentials is 
given as 
\begin{eqnarray}
\label{potens}
\phi_1 (r) &=& - \frac{1}{2 \left(r - 1 \right)} \nonumber \\
\phi_2 (r) &=& - \frac{1}{2r} \left[ 1 - \frac{3}{2r} + 12 
\left( \frac{1}{2r} \right)^2 \right ] \nonumber \\
\phi_3 (r) &=& -1 + \left(1 - \frac{1}{r} \right)^{1/2} \nonumber \\
\phi_4 (r) &=& \frac{1}{2} \ln \left(1 - \frac{1}{r} \right)
\end{eqnarray}
in all of which, the length of the radial coordinate, $r$, has been 
scaled in units of Schwarzschild radius, defined as 
$r_g = 2GM_{\mathrm{BH}}/c^2$ (with $M_{\mathrm{BH}}$ being the mass
of the black hole, $G$ the universal gravitational constant and $c$ 
the velocity of light in vacuum). At various points of time, each of
these potentials has been introduced in accretion literature
by~\citet{pw80} for 
$\phi_1$,~\citet{nw91} for $\phi_2$, and~\citet{abn96} for $\phi_3$
and $\phi_4$, respectively. A comparative overview of the  
physical properties of these potentials has been given 
by~\citet{ds01},~\citet{das02} and~\citet{dpm03}. 

\begin{figure}
\vskip -4.0cm
\begin{center}
\includegraphics[scale=0.62, angle=0]{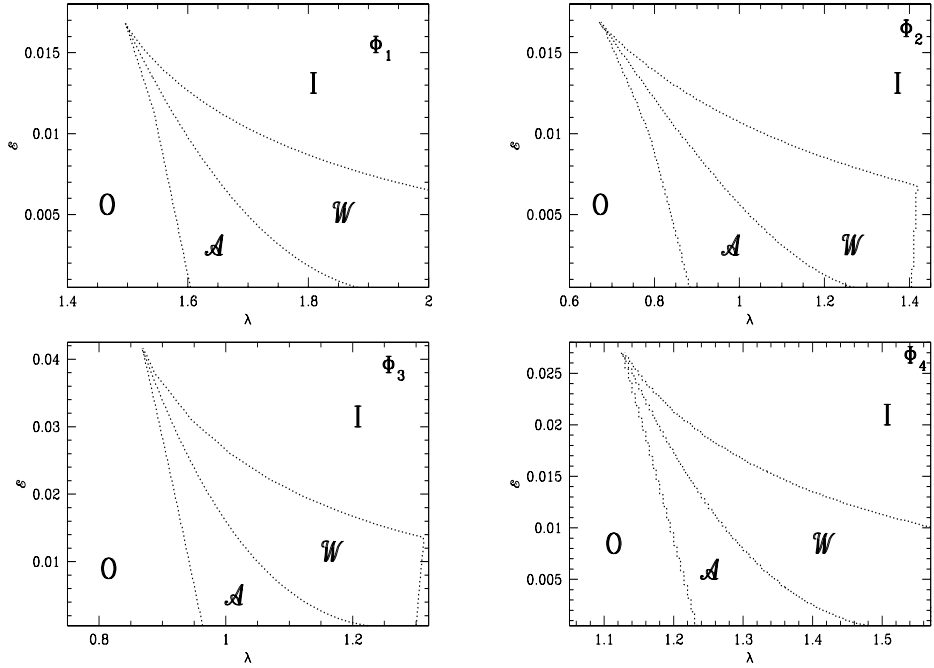}
\vskip -5.0cm
\caption{\label{f1} \small{Regions of multi-transonicity for both
accretion and wind in the parameter space of $\mathcal E$ and 
$\lambda$, under four different pseudo-Newtonian potentials.}}
\end{center}
\end{figure}

Considering polytropic flows first, Fig.~\ref{f1} gives the
plots of $\mathcal E$ versus $\lambda$ parameter space for all
the potentials $\phi_i (r)$. The region marked $\mathbf O$ gives
the values of $\mathcal E$ and $\lambda$ corresponding to the
single outer critical point. Similarly $\mathbf I$ indicates
the region of the parameter space that will yield the lone inner
critical point.
Multi-transonicity is to be obtained for the values of $\mathcal E$
and $\lambda$ contained within the wedge-shaped region, which is
further subdivided into two regions, $\mathcal A$ (for accretion)
and $\mathcal W$ (for wind). The area of multi-transonicity shifts
according to the choice of a particular potential.

\begin{figure}
\vskip -2.0cm 
\begin{center}
\includegraphics[scale=0.62, angle=0]{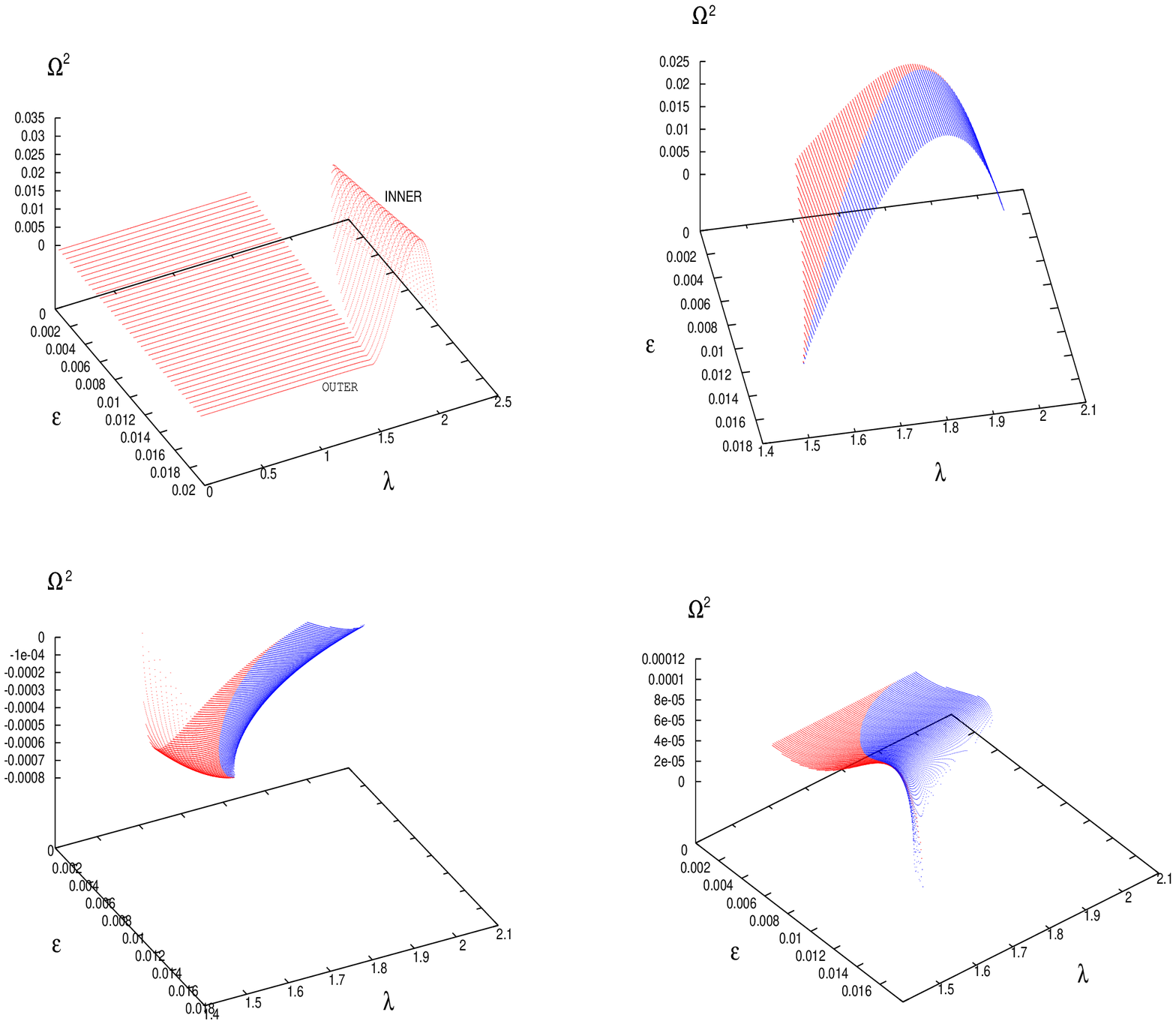}
\vskip -5.0cm 
\caption{\label{f2} \small{Variation of $\Omega^2$
with respect to $\mathcal E$ and $\lambda$, under the pseudo-Newtonian 
potential $\phi_1$. The plot in the top left corner gives the 
variation for the regions of a single critical point only. The 
remaining three figures are for the multi-transonic region. Clockwise
from the top right corner will give the variations for the inner
critical point, the outer critical point and the middle critical 
point, respectively.}}
\end{center}
\end{figure}

\begin{figure}
\vskip -4.0cm
\begin{center}
\includegraphics[scale=0.60, angle=0]{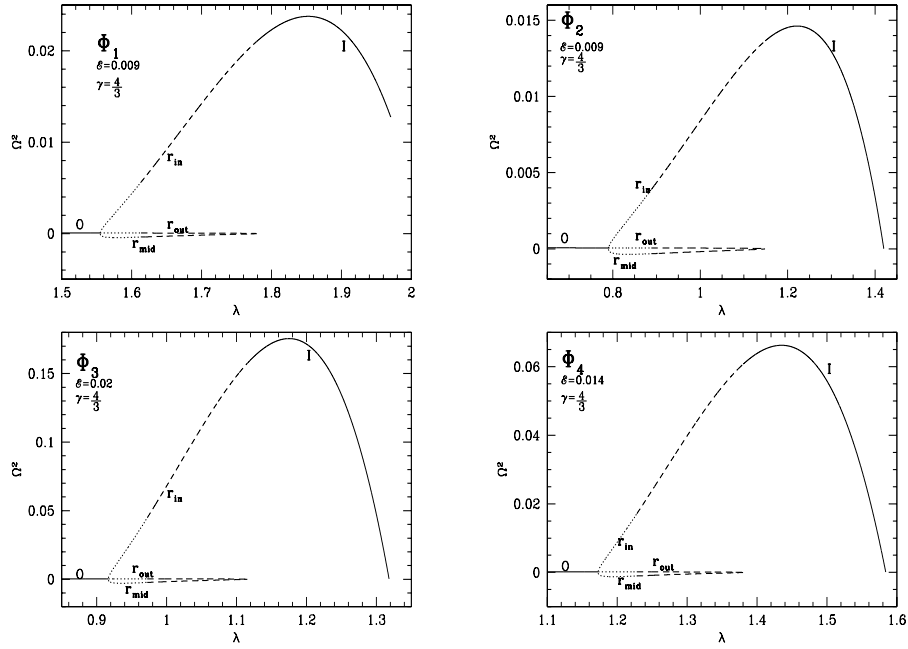}
\vskip -5.0cm
\caption{\label{f3} \small{Variation of $\Omega^2$
with respect to $\lambda$, for fixed values
of $\mathcal E$, in a polytropic flow. Each plot shows the 
birth of two critical
points at a given value of $\lambda$, which coalesce later
at a higher value of $\lambda$.}}
\end{center}
\end{figure}

\begin{figure}
\vskip -4.0cm
\begin{center}
\includegraphics[scale=0.60, angle=0]{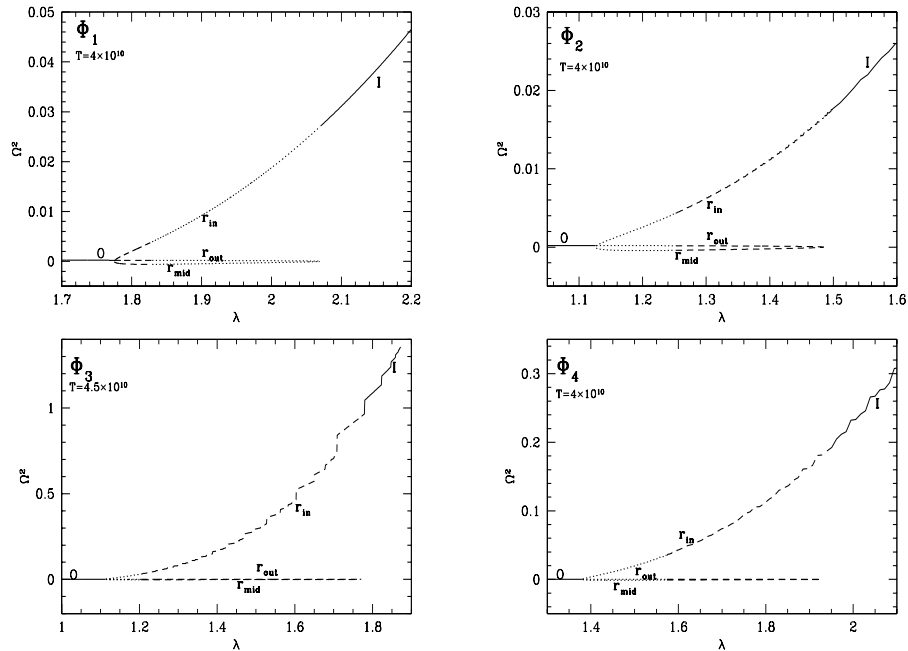}
\vskip -5.0cm
\caption{\label{f4} \small{Variation of $\Omega^2$ with respect 
to $\lambda$, for fixed values of $T$, in an isothermal flow. 
Unlike the case of polytropic flows, here the eigenvalues related 
to the inner critical point (a saddle point) continue to grow 
monotonically beyond the range of multi-transonicity.}}
\end{center}
\end{figure}

For the~\citet{pw80} potential, $\phi_1$, 
the variation of $\Omega^2$ against $\mathcal E$ and $\lambda$ is 
given in Fig.~\ref{f2}. The plot in the top left corner of this 
panel gives the dependence of $\Omega^2$ on $\mathcal E$ and 
$\lambda$ for the regions of the single critical point. The 
flat portion is related to the outer critical point, while the
hump is related to the inner critical point. Both of these points
are saddle points. The vacant wedge-shaped area between these two
regions is the region of multi-transonicity, which has been 
separately represented for the behaviour of the three distinct 
critical points in the remaining three plots in the panel. The 
top right corner shows the variation of the inner critical point,
which is a saddle point. The bottom left plot gives the variation
of the eigenvalues for the centre-type middle critical point, 
while the properties of the outermost saddle-type critical point
are given by the bottom right plot in Fig.~\ref{f2}.

A common feature of all the three separate plots for the 
eigenvalues of the multi-transonic region is that they contain
information on both accretion and wind. Here the lightly shaded 
regions (coloured red in the online version) in the plots indicate 
accretion. The criterion to 
distinguish accretion from wind has been that the flow rate 
through the inner critical point (a saddle point), 
$\dot{\mathcal{M}}_{\mathrm{in}}$, has to be greater than the 
corresponding flow rate through the outer critical point (again 
a saddle point), $\dot{\mathcal{M}}_{\mathrm{out}}$. The exact 
reversal of this requirement will give the wind region, which
has been given by the darkly shaded portion (coloured blue in 
the online version) of each plot. These
points can be better appreciated from the colour images on the 
archive version of this paper.

Another property that one could discern from the surfaces in 
the $(\Omega^2, {\mathcal E}, \lambda )$ space is that the 
surface pertaining to the centre-type point has a minimum and 
then it rises, while the surface for the second saddle point 
has a miximum and then it dips. So the two of them taken
together, they coalesce and ``annihilate" each other at a 
large value of the angular momentum, and then all that is 
left behind is a single saddle point for the high angular 
momentum region of the parameter space. 

This contention is very illustratively borne out by the top
left plot of Fig.~\ref{f3}, in which $\Omega^2$ for all the
three critical points has been plotted against $\lambda$, for 
a suitably chosen value of $\mathcal E$, in the case of a flow 
driven by the~\citet{pw80} potential. Initially there is region 
of a single saddle point, followed by the birth of a centre-type 
point and another saddle point. Physically this is what it has 
to be, because with a centre-type point but without another 
saddle point, the flow solutions will all curl about the 
centre-type point, and there will be no means of connecting
the event horizon with infinity through a solution. To avoid
this, and to make accretion a feasible proposition, the 
existence of the inner saddle point should be most crucial. 
The other three plots in Fig.~\ref{f3} demonstrate this same
feature for the other pseudo-Newtonian potentials. 

The whole situation is qualitatively similar for isothermal
flows. The case for multi-transonicity in the parameter space
of $T$ and $\lambda$ has been discussed in detail by~\citet{dpm03}.
It is a simple extension from here to study the dependence of
$\Omega^2$ on $T$ and $\lambda$. This has been shown in 
Fig.~\ref{f4}, which gives the variation of $\Omega^2$ against
$\lambda$, for fixed values of $T$. Once again it is evident 
that the multi-transonic region begins and ends, respectively,
with the production and annihilation of a pair of critical 
points --- one a saddle and another a centre-type point. A
crucial difference for isothermal flows, however, is that beyond 
the multi-transonic region, for high values of angular momentum, 
the eigenvalues in relation to the lone saddle point increase 
monotonically in magnitude. 

A few points are worth emphasising upon once again at the end of this
whole discussion, dwelling on both kinds of flows --- polytropic
and isothermal. Without encountering the difficulties of 
numerical integration, much predictive insight has been derived
about the qualitative character of the flow, not the least of 
which, crucially, is the exclusive nature of the critical 
points, and in relation to that, important physical features 
of multi-transonicity itself. 

\section{Time-dependent stability analysis of stationary solutions}
\label{sec5}

Under the condition of hydrostatic equilibrium in the vertical direction,
the time-dependent generalisation of the governing equations for an
axisymmetric pseudo-Schwarzschild disc is given as 
\begin{equation}
\label{dyneuler}
\frac{\prt v}{\prt t} + v \frac{\prt v}{\prt r}
+ \frac{1}{\rho} \frac{\prt P}{\prt r} + \phi^{\prime}(r)
-\frac{\lambda^2}{r^3}=0
\end{equation}
and 
\begin{equation}
\label{surden}
\frac{\prt \Sigma}{\prt t}+ \frac{1}{r} \frac{\prt}{\prt r}
\left( \Sigma vr \right)=0
\end{equation}
in which the surface density of the thin disc $\Sigma$, is to be
expressed as $\Sigma \cong \rho H$~\citep{fkr02}. 
Making use of Eq.(\ref{aitchpol}) 
and the polytropic relation $P = K \rho^{\gamma}$, Eq.(\ref{surden}) 
can be rendered as 
\begin{equation}
\label{volden}
\frac{\prt}{\prt t} \left[\rho^{(\gamma +1)/2}\right]
+ \frac{\sqrt{\phi^{\prime}}}{r^{3/2}}
\frac{\prt}{\prt r} \left[ \rho^{(\gamma +1)/2} v
\frac{r^{3/2}}{\sqrt{\phi^{\prime}}} \right] =0
\end{equation}

Defining a new variable 
$f=\rho^{(\gamma +1)/2} v r^{3/2}/\sqrt{\phi^{\prime}}$, it is quite
obvious from the form of Eq.(\ref{volden}) that the stationary value 
of $f$ will be a constant, $f_0$, which can be closely identified with
the matter flux rate. This follows a similar approach to spherically 
symmetric flows made by~\citet{pso80} and~\citet{td92}. For an 
axisymmetric disc, which has no dependence on any angle variable, 
this approach has also been adopted for a flow driven simply by a 
Newtonian potential~\citep{ray03}. The present treatment, of course,
is of a more general nature, the disc flow being driven by a general
pseudo-Newtonian potential, $\phi (r)$. In this system, a perturbation 
prescription of the form $v(r,t) = v_0(r) + v^{\prime}(r,t)$ and 
$\rho (r,t) = \rho_0 (r) + \rho^{\prime}(r,t)$, will give, on 
linearising in the primed quantities,  
\begin{equation}
\label{effprime}
f^{\prime} = f_0 \left[\left( \frac{\gamma +1}{2} \right)
\frac{\rho^{\prime}}
{\rho_0} + \frac{v^{\prime}}{v_0} \right]
\end{equation}
with the subscript $0$ denoting stationary values in all cases. From
Eq.(\ref{volden}), it then becomes possible to set down the density
fluctuations $\rho^{\prime}$, in terms of $f^{\prime}$ as 
\begin{equation}
\label{flucden}
\frac{\prt \rho^{\prime}}{\prt t} + \beta^2
\frac{v_0 \rho_0}{f_0} \left(\frac{\prt f^{\prime}}{\prt r}\right)=0
\end{equation}
with $\beta^2 = 2(\gamma +1)^{-1}$, as before. 
Combining Eqs.(\ref{effprime}) and (\ref{flucden}) will then render
the velocity fluctuations as 
\begin{equation}
\label{flucvel}
\frac{\prt v^{\prime}}{\prt t}= \frac{v_0}{f_0}
\left(\frac{\prt f^{\prime}}{\prt t}+{v_0}
\frac{\prt f^{\prime}}{\prt r}\right)
\end{equation}
which, upon a further partial differentiation with respect to time,
will give 
\begin{equation}
\label{flucvelder2}
\frac{{\prt}^2 v^{\prime}}{\prt t^2}=\frac{\prt}{\prt t} \left[
\frac{v_0}{f_0} \left(\frac{\prt f^{\prime}}{\prt t}\right) \right]
+ \frac{\prt}{\prt t} \left[ \frac{v_0^2}{f_0} \left(
\frac{\prt f^{\prime}}{\prt r}\right) \right]
\end{equation}

From Eq.(\ref{dyneuler}) the linearised fluctuating part could be
extracted as 
\begin{equation}
\label{fluceuler}
\frac{\prt v^{\prime}}{\prt t}+ \frac{\prt}{\prt r}
\left( v_0 v^{\prime} + c_{\mathrm{s0}}^2 
\frac{\rho^{\prime}}{\rho_0}\right) =0
\end{equation}
with $c_{\mathrm{s0}}$ being the speed of sound in the steady state.
Differentiating Eq.(\ref{fluceuler}) partially with respect to $t$,
and making use of Eqs.(\ref{flucden}), (\ref{flucvel}) and  
(\ref{flucvelder2}) to substitute for all the first and second-order
derivatives of $v^{\prime}$ and $\rho^{\prime}$, will deliver the result 
\begin{equation}
\label{interm}
\frac{\prt}{\prt t} \left[\frac{v_0}{f_0}
\left( \frac{\prt f^{\prime}}{\prt t}\right)\right]
+ \frac{\prt}{\prt t} \left[\frac{v_0^2}{f_0}
\left( \frac{\prt f^{\prime}}{\prt r}\right)\right]
+ \frac{\prt}{\prt r} \left[\frac{v_0^2}{f_0}
\left( \frac{\prt f^{\prime}}{\prt t}\right)\right]
+ \frac{\prt}{\prt r} \left[\frac{v_0}{f_0}
\left(v_0^2 - \beta^2 c_{\mathrm{s0}}^2 \right)
\frac{\prt f^{\prime}}{\prt r}\right] = 0
\end{equation}
all of whose terms can be ultimately rendered into a compact
formulation that looks like
\begin{equation}
\label{compact}
\prt_\mu \left( {\mathrm{f}}^{\mu \nu} \prt_\nu
f^{\prime}\right) = 0
\end{equation}
in which the Greek indices are made to run from $0$ to $1$, with 
the identification that $0$ stands for $t$, and $1$ stands for $r$.
An inspection of the terms in the left hand side of Eq.(\ref{interm})
will then allow for constructing the symmetric matrix
\begin{equation}
\label{matrix}
{\mathrm{f}}^{\mu \nu } = \frac{v_0}{f_0}
\pmatrix
{1 & v_0 \cr 
v_0 & v_0^2 - \beta^2 c_{\mathrm{s0}}^2}
\end{equation}
Now the d'Alembertian for a scalar in curved space is given in terms
of the metric ${\mathrm{g}}_{\mu \nu}$ by~\citep{vis98}
\begin{equation}
\label{alem}
\Delta \psi \equiv \frac{1}{\sqrt{-\mathrm{g}}}
\prt_\mu \left({\sqrt{-\mathrm{g}}}\, {\mathrm{g}}^{\mu \nu} \prt_\nu
\psi \right)
\end{equation}
with $\mathrm{g}^{\mu \nu}$ being the inverse of the matrix implied
by ${\mathrm{g}}_{\mu \nu}$. Using the equivalence that
${\mathrm{f}}^{\mu \nu } = \sqrt{-\mathrm{g}}\, {\mathrm{g}}^{\mu \nu}$,
and therefore $\mathrm{g} = \det \left({\mathrm{f}}^{\mu \nu }\right)$,
it is immediately possible to set down an effective metric for the 
propagation of an acoustic disturbance as
\begin{equation}
\label{metric}
\mathrm{g}^{\mu \nu}_{\mathrm{eff}} =
\pmatrix
{1 & v_0 \cr
v_0 & v_0^2 - \beta^2 c_{\mathrm{s0}}^2}
\end{equation}
which can be shown to be entirely identical to the metric of a wave 
equation for a scalar field in curved space-time, obtained through a
somewhat different approach~\citep{vis98}. The inverse effective
metric, $\mathrm{g}_{\mu \nu}^{\mathrm{eff}}$, can be easily derived
by inversion of the matrix given in Eq.(\ref{metric}), and this will 
give $v_0^2 = \beta^2 c_{\mathrm{s0}}^2$ as the horizon condition of an
acoustic black hole for inflow solutions~\citep{vis98}. 

A little readjustment of terms in Eq.(\ref{interm}) will finally give 
an equation for the perturbation as  
\begin{equation}
\label{tpert}
\frac{{\prt}^2 f^{\prime}}{\prt t^2} +2 \frac{\prt}{\prt r}
\left(v_0 \frac{\prt f^{\prime}}{\prt t} \right) + \frac{1}{v_0}
\frac{\prt}{\prt r}\left[ v_0 \left(v_0^2- 
\beta^2 c_{\mathrm{s0}}^2 \right)
\frac{\prt f^{\prime}}{\prt r}\right] = 0
\end{equation}
whose form, as can be easily seen, {\em remains
invariant under any possible choice of a pseudo-potential}. This is
entirely to be expected since the potential, being independent of 
time, will have its explicit presence in only the stationary 
background flow. Once again it is a simple exercise to check that
an expression with the same form as Eq.(\ref{tpert}) could be derived 
for the case of isothermal flows, but which will, however, have 
$\beta =1$ and $c_{\mathrm{s0}}$ as a constant.  

The perturbation is now fashioned to behave like a travelling wave,
whose wavelength is constrained to remain small, i.e. it is to be 
smaller than any characteristic length scale in the system. This 
kind of a treatment has been carried out before on spherically 
symmetric flows~\citep{pso80} and on axisymmetric flows driven by
a Newtonian potential~\citep{ray03}. In both these cases the radius
of the accretor was chosen as the characteristic length scale in 
question, and the wavelength of the perturbation was chosen to be
much smaller than this length scale. In the present study, which 
is devoted to an axisymmetric flow driven on to a black hole by a 
pseudo-Newtonian potential,
the radius of the event horizon could be a choice for such a length 
scale. As a result, the frequency, $\omega$, of the waves should be 
large. A solution of the kind $f^{\prime}(r,t) = g_\omega (r) 
\exp(-{\mathrm i} \omega t)$ is used in Eq.(\ref{tpert}), to 
give the expression 
\begin{equation}
\label{gee}
\left(v_0^2 - \beta^2 c_{\mathrm{s0}}^2\right) 
\frac{{\mathrm d^2} g_\omega}{{\mathrm d}r^2} +\left[ 3 v_0 
\frac{{\mathrm d}v_0}{{\mathrm d}r} - \frac{1}{v_0}
{\frac{\mathrm d}{{\mathrm d}r}} \left( v_0 \beta^2 c_{\mathrm{s0}}^2 \right) 
- 2{\mathrm{i}} \omega v_0 \right] \frac{{\mathrm d} g_\omega}{{\mathrm d}r}
-\left( 2{\mathrm i} \omega \frac{{\mathrm d}v_0}{{\mathrm d}r} 
+ \omega^2 \right)g_\omega = 0
\end{equation}
beyond which stage, mindful of the constraint that $\omega$ is large,
the spatial part of the perturbation, $g_\omega (r)$, is first prescribed 
to be a power series in the form
\begin{equation}
\label{pow}
g_\omega (r)= \exp \left[\sum_{n=-1}^{\infty} 
\frac{k_n(r)}{\omega^n} \right]
\end{equation}
and this is then used in Eq.(\ref{gee}). Following this, the three successive
highest order terms in $\omega$ will be obtained as $\omega^2$, $\omega$ 
and $\omega^0$. Collecting all the coefficients of each of these terms,
summing these coefficients up, and then separately setting each individual 
sum to zero, will yield, respectively, for $\omega^2$, $\omega$ 
and $\omega^0$, the conditions
\begin{equation}
\label{omegsq}
\left(v_0^2 - \beta^2 c_{\mathrm{s0}}^2\right)
\left( \frac{{\mathrm d}k_{-1}}{{\mathrm d}r} \right)^2
-2 {\mathrm i} v_0 \frac{{\mathrm d}k_{-1}}{{\mathrm d}r} -1 = 0
\end{equation}
\begin{equation}
\label{omeg1}
\left(v_0^2 - \beta^2 c_{\mathrm{s0}}^2\right)
\left( \frac{{\mathrm d}^2 k_{-1}}{{\mathrm d}r^2}
+ 2 \frac{{\mathrm d}k_{-1}}{{\mathrm d}r} 
\frac{{\mathrm d}k_0}{{\mathrm d}r} \right)
+ \left[ 3 v_0 \frac{{\mathrm d}v_0}{{\mathrm d}r} 
- \frac{1}{v_0}\frac{\mathrm d}{{\mathrm d}r}
\left( v_0 \beta^2 c_{\mathrm{s0}}^2 \right) \right]
\frac{{\mathrm d}k_{-1}}{{\mathrm d}r}
- 2{\mathrm i} v_0 \frac{{\mathrm d}k_0}{{\mathrm d}r} 
- 2{\mathrm i} \frac{{\mathrm d}v_0}{{\mathrm d}r} = 0
\end{equation}
and
\begin{equation}
\label{omeg0}
\left( v_0^2 - \beta^2 c_{\mathrm{s0}}^2 \right)
\left[ \frac{{\mathrm d}^2 k_0}{{\mathrm d}r^2}
+ 2 \frac{{\mathrm d}k_{-1}}{{\mathrm d}r} 
\frac{{\mathrm d}k_1}{{\mathrm d}r} + 
\left( \frac{{\mathrm d}k_0}{{\mathrm d}r} \right)^2
\right] + \left [3 v_0 \frac{{\mathrm d}v_0}{{\mathrm d}r}
- \frac{1}{v_0} \frac{\mathrm d}{{\mathrm d}r}
\left( v_0 \beta^2 c_{\mathrm{s0}}^2 \right) \right]
\frac{{\mathrm d}k_0}{{\mathrm d}r} - 
2{\mathrm{i}}{v_0}\frac{{\mathrm d}k_1}{{\mathrm d}r} = 0
\end{equation}
Out of these, the first two, i.e. Eqs.(\ref{omegsq}) and (\ref{omeg1}),
will deliver the solutions
\begin{equation}
\label{kayminus1}
k_{-1} = \int \frac{{\mathrm{i}}}{v_0 \pm \beta c_{\mathrm{s0}}} \,
{\mathrm d}r
\end{equation}
and
\begin{equation}
\label{kaynot}
k_0 = - \frac{1}{2} \ln \left( \beta v_0 c_{\mathrm{s0}} \right) 
+ {\mathrm{constant}}
\end{equation}
respectively. The two expressions above give the leading terms in 
the power series of $g_\omega (r)$. For self-consistency it will 
be necessary to show that successive terms follow the condition
$\omega^{-n}\vert k_n(r)\vert \gg \omega^{-(n+1)}\vert k_{n+1}(r)\vert$,
i.e. the power series given by $g_\omega (r)$ converges rapidly as 
$n$ increases. This requirement can be shown to be very much true, 
if the behaviour of the first three terms in $k_n(r)$ is any
indication. From Eqs.~(\ref{kayminus1}), (\ref{kaynot}) 
and (\ref{omeg0}), respectively, these terms can be shown to go 
asymptotically as 
$k_{-1} \sim r$, $k_0 \sim \ln r$ and $k_1 \sim r^{-1}$, given the
condition that $v_0 \sim r^{-5/2}$ on large length scales, while 
$c_{\mathrm{s0}}$ approaches its constant ambient value. Under the
regime of high-frequency travelling waves, this will imply 
$\omega \vert k_{-1} \vert \gg \vert k_0 \vert \gg \omega^{-1} 
\vert k_1 \vert$, and, therefore, it should suffice here entirely 
to consider the two leading terms only, as given by Eqs.(\ref{kayminus1}) 
and (\ref{kaynot}), in the power series expansion 
of $g_\omega (r)$, as Eq.~(\ref{pow}) gives it. So with the help 
of these two terms, one may then set down an expression for 
the perturbation as
\begin{equation}
\label{fpertur}
f^{\prime}(r,t) \simeq \frac{A_\pm}
{\sqrt{\beta v_0 c_{\mathrm{s0}}}}e^{-{\mathrm i} \omega t}
\exp \left ( \int  \frac{{\mathrm i} \omega}{v_0 \pm \beta 
c_{\mathrm{s0}}} \, {\mathrm d}r \right)
\end{equation}
which should be seen as a linear superposition
of two waves with arbitrary constants $A_+$ and $A_-$. Both these two 
waves move with velocity $\beta c_{\mathrm{s0}}$ relative to the fluid,
one against the bulk flow and the other along with it, while the bulk
flow itself has velocity $v_0$. With the help of Eq.(\ref{flucden}) it 
should then be easy to express the density fluctuations in terms of
$f^{\prime}$ as 
\begin{equation}
\label{effden}
\rho^{\prime} \simeq \beta^2 \frac{v_0 \rho_0}{f_0}
\left(\frac{f^{\prime}}{v_0 \pm \beta c_{\mathrm{s0}}}\right)
\end{equation}
and as a further step, with the help of Eq.(\ref{effprime}), it should
also be possible to set down the velocity fluctuations as 
\begin{equation}
\label{effvel}
v^{\prime} \simeq \pm \beta \frac{v_0 c_{\mathrm{s0}}}{f_0}
\left(\frac{f^{\prime}}{v_0 \pm \beta c_{\mathrm{s0}}}\right)
\end{equation}
with the positive and the negative signs indicating incoming and 
outgoing waves, respectively. 

In a unit volume of the fluid, the kinetic energy content is
\begin{equation}
\label{ekin}
{\mathcal E}_{\mathrm{kin}} = \frac{1}{2} \left(\rho_0
+ \rho^{\prime}\right)\left(v_0 + v^{\prime}\right)^2
\end{equation}
while the potential energy per unit volume of the fluid is the sum of 
the gravitational energy, the rotational energy and the internal energy,
and is given by
\begin{equation}
\label{epot}
{\mathcal E}_{\mathrm{pot}} = \left(\rho_0 + \rho^{\prime}\right) \phi
- \left(\rho_0 + \rho^{\prime}\right) \frac{\lambda^2}{2r^2}
+ \rho_0 \epsilon + \rho^{\prime} \frac{\prt}{\prt \rho_0}
\left(\rho_0 \epsilon \right) + \frac{1}{2}{\rho^{\prime}}^2
\frac{{\prt}^2}{\prt \rho_0^2}\left(\rho_0 \epsilon \right)
\end{equation}
where $\epsilon$ is the internal energy per unit mass~\citep{ll87}. 
In Eqs.(\ref{ekin}) and (\ref{epot}), the first-order terms vanish on 
time-averaging. In that case the leading contribution to the total 
energy in the 
perturbation comes from the second-order terms, which are given by
\begin{equation}
\label{order2}
{\mathcal E}_{\mathrm{pert}} = \frac{1}{2} \rho_0 {v^{\prime}}^2
+ v_0 \rho^{\prime} v^{\prime} +  \frac{1}{2} {\rho^{\prime}}^2
\frac{{\prt}^2}{\prt \rho_0^2} \left(\rho_0 \epsilon \right)
\end{equation}

If the perturbation is constrained to be adiabatic, then the 
condition ${\mathrm d}s=0$ can be imposed on the thermodynamic relation, 
${\mathrm d} \epsilon = T{\mathrm d}s + 
\left( P/\rho^2 \right) {\mathrm d} \rho$, which will then give
\begin{equation}
\label{adiab}
\frac{{\prt}^2}{\prt \rho_0^2} \left(\rho_0 \epsilon \right)
{\bigg{\vert}}_s = \frac{c_{\mathrm{s0}}^2}{\rho_0}
\end{equation}
Combining the results from Eqs.(\ref{fpertur}), (\ref{effden}), 
(\ref{effvel}), (\ref{order2}) and (\ref{adiab}), and upon 
time-averaging over the square of the fluctuations, will give
\begin{equation}
\label{epertfin}
{\mathcal E}_{\mathrm{pert}} \simeq \frac{1}{2} 
\frac{\beta^2 A_{\pm}^2}{f_0^2}
\frac{v_0 \rho_0}{\left(v_0 \pm \beta c_{{\mathrm s0}}\right)^2}
\left[\frac{\beta c_{\mathrm{s0}}}{2} 
\left(1 + \frac{1}{\beta^2} \right) \pm v_0 \right]
\end{equation}
with the factor $1/2$ in Eq.(\ref{epertfin}) arising from the 
time-averaging of the fluctuations.

The total energy flux in the perturbation is obtained by multiplying
${\mathcal E}_{\mathrm{pert}}$
by the propagation velocity $(v_0 \pm \beta c_{\mathrm{s0}})$ and then 
by integrating over the area of the cylindrical face of the accretion
disc, which is $2 \pi rH$. Here $H$ is to be substituted from 
Eq.(\ref{aitchpol}), and together with the condition that $H \ll r$ 
in the thin-disc approximation, and the fact that the speed of sound 
has a functional dependence on the density going as 
$c_{\mathrm{s0}}^2 = \gamma K \rho_0^{\gamma -1}$,  
an expression for the energy flux will be obtained as
\begin{equation}
\label{flux}
{\mathcal F} = \frac{\pi A_{\pm}^2 \sqrt{K} \beta^2}{f_0}
\left[\pm 1 + \frac{1- \beta^2}{2 \beta
\left( M \pm \beta \right)}\right]
\end{equation}
where $ M \equiv M(r) = v_0/c_{\mathrm{s0}}$, is the Mach number, which
under a generic outer boundary condition, tends to vanishingly small
values on large length scales of the accretion disc. All other factors
in the expression of $\mathcal F$ are constants. It is quite evident 
once again that the forms of both ${\mathcal E}_{\mathrm{pert}}$ and 
${\mathcal F}$ are invariant under the choice of any pseudo-potential, 
and therefore all arguments for the stability of the flow will be valid 
for any choice. Indeed, this should be especially true on large length 
scales of the flow, where all $\phi (r)$ will asymptotically tend to 
the Newtonian limit. 

As regards the question of the behaviour of the perturbation, it can
be seen that for sub-critical solutions, the condition $M(r) < \beta$
always holds everywhere for both the incoming and the outgoing modes
of the perturbation, which, of course, implies that the total energy 
in the wave remains finite as the wave propagates. The perturbation
manifests no unbridled divergence anywhere, and all sub-critical 
solutions are, therefore, stable. 

For the critical solution, it is evident that with the choice of the 
upper sign in Eq.(\ref{flux}), i.e. for incoming waves, which propagate 
along with the bulk flow, once again there is no growth behaviour for 
the perturbation to be seen anywhere and all stationary inflow solutions 
are, therefore, stable for this propagating mode of the perturbation. 

For the choice of the lower sign, i.e. for outwardly propagating 
radial waves, it is first of all very obvious 
from Eq.(\ref{flux}), that there is no unrestrained growth of the energy 
flux of the perturbation in the region where $M(r) < \beta$, and so 
the disc system remains stable in this sub-critical region.
The case for stability is even better argued in the region of the flow 
where $M(r) > \beta$. In this region, which has a finite spatial extent, 
all outwardly propagating waves will be swept away in finite time by 
the inward bulk flow right up to the event horizon of the black hole, 
and this physical effect will also ensure the stability of the flow in 
this region. This very same line of reasoning was adopted by~\citet{gar79}
to argue for the stability of the critical solution in the supersonic
region of a spherically symmetric flow. 
As an interesting aside, it may also be mentioned here that
this argument is well in consonance with the analogy of an acoustic 
black hole. Any disturbance within the sonic horizon for inflows, will
not be able to propagate outwards through the horizon. Therefore, it 
can be established generally that in an inviscid pseudo-Schwarzschild 
system, all physically well-behaved inflow solutions will be stable 
under the influence of a linearised high-frequency radial perturbation, 
propagating either inwardly or outwardly. 

\section{Concluding remarks}
\label{sec6}

Saddle points are inherently unstable, and to make a solution pass
through such a point, after starting from an outer boundary condition,
will entail an infinitely precise fine-tuning of that boundary 
condition~\citep{rb02}. Nevertheless, transonicity, i.e. generating 
a solution
through a saddle point, is not a matter of doubt. The key to this 
paradox lies in considering a time-dependent evolution of the flow. 
In this paper, time-dependence in the flow has indeed been taken up,
although in a perturbative manner, which has shown that all solutions
of physical interest are stable. So the selection of a solution 
through a saddle point will have to fall back on non-perturbative
arguments~\citep{rb02}. This is a relatively simple proposition for 
the case of
a flow in the Newtonian framework. The difficulties against this 
line of attack, however, are understandably much greater when the
flow properties are studied in a general relativistic framework. 
Notwithstanding this fact, as a preliminary foray into this question, 
it should be important to have some understanding of the behaviour 
of the general relativistic flow solutions in their stationary phase 
portrait. This will be a matter of a subsequent investigation, for 
flows both on to a Schwarzschild black hole and a Kerr black hole. 

\section*{Acknowledgements}

This research has made use of NASA's Astrophysics Data System. 
S. Chaudhury would like to acknowledge the kind hospitality 
provided by HRI, Allahabad, India, under a visiting student
research programme. A. K. Ray expresses his indebtedness to 
J. K. Bhattacharjee for providing much helpful insight. The 
work of T. K. Das has been partially supported by a grant 
(No. 94-2752-M-007-001-PAE) provided by TIARA, Taiwan.

\end{document}